\documentclass[aps,prl,twocolumn,showpacs,superscriptaddress,groupedaddress]{revtex4}
\usepackage[utf8]{inputenc} 
\usepackage{mathtools,amssymb,lipsum, nccmath}
\usepackage{bm}

\usepackage{graphicx} 
\usepackage{color}
\usepackage{float}

\usepackage{booktabs} 
\usepackage{array} 
\usepackage{paralist} 
\usepackage{verbatim} 
\usepackage{subfig} 
\usepackage{hyperref}

\makeatletter
\newcommand*{\rom}[1]{\expandafter\@slowromancap\romannumeral #1@}
\makeatother

\begin{document}

\title{Floquet Nonadiabatic Nuclear Dynamics with Photoinduced Lorentz-Like Force in Quantum Transport}
\author{Jingqi Chen}
\affiliation{Fudan University, Shanghai 200433, China}
\affiliation{Department of Chemistry, School of Science, Westlake University, Hangzhou, Zhejiang 310024, China}
\affiliation{Institute of Natural Sciences, Westlake Institute for Advanced Study, Hangzhou, Zhejiang 310024, China}

\author{Wei Liu}
\affiliation{Department of Chemistry, School of Science, Westlake University, Hangzhou, Zhejiang 310024, China}
\affiliation{Institute of Natural Sciences, Westlake Institute for Advanced Study, Hangzhou, Zhejiang 310024, China}

\author{Wenjie Dou}
\email{douwenjie@westlake.edu.cn} 
\affiliation{Department of Chemistry, School of Science, Westlake University, Hangzhou, Zhejiang 310024, China}
\affiliation{Department of Physics, School of Science, Westlake University, Hangzhou, Zhejiang 310024, China}
\affiliation{Institute of Natural Sciences, Westlake Institute for Advanced Study, Hangzhou, Zhejiang 310024, China}

\begin{abstract}
In our recent paper [Mosallanejad $et$ $al.$, Phys. Rev. B 107(18), 184314, 2023], we have derived a Floquet electronic friction model to describe nonadiabatic molecular dynamics near metal surfaces in the presence of periodic driving. In this work, we demonstrate that Floquet driving can introduce an anti-symmetric electronic friction tensor in quantum transport, resulting in circular motion of the nuclei in the long-time limit. Furthermore, we show that such a Lorentz-like force strongly affects nuclear motion: at lower voltage bias, Floquet driving can increase the temperature of nuclei; at larger voltage bias, Floquet driving can decrease the temperature of nuclei. In addition, Floquet driving can affect electron transport strenuously. Finally, we show that there is an optimal frequency that maximizes electron current. We expect that the Floquet electronic friction model is a powerful tool to study nonadiabatic molecular dynamics near metal surfaces under Floquet driving in complex systems.

\end{abstract}
 
\maketitle

\section{Introduction}
The Born-Oppenheimer approximation has long stood as a pillar in quantum chemistry, offering a powerful framework to decouple electronic and nuclear motions in molecular systems. However, the Born-Oppenheimer approximation can break down strenuously, in particular when dealing with a manifold of electronic states. One such scenario is the scattering process on metal surfaces, where the conventional adiabatic separation of electronic and nuclear degrees of freedom becomes inadequate \cite{jasper2004introductory,wodtke2004electronically,yarkony2012nonadiabatic}. The electronic friction (EF) model serves as the first order correction to the Born-Oppenheimer approximation, offering a powerful tool to study nonadiabatic dynamics near metal surfaces \cite{head1995molecular, dou2018perspective, dou2017born, dou2017universality,dou2018universality,maurer2017mode, lu2010blowing, bode2012current}. 
Similar to the Nakajima-Zwanzig formulation, in the EF model, the fast quantum electronic motion serves as environment, resulting in both frictional force and random force on the slow and classical nuclear motion.

Within the EF model, a notable distinction arises in the treatment of the friction tensor. This friction tensor, a key concept in electronic friction theory, can exhibit both symmetric and anti-symmetric components. The symmetric part of the friction tensor introduces relaxation on nuclear motion, which has been widely used to study energy relaxation in surface scattering processes \cite{huang2000vibrational, bunermann2015electron,maurer2017mode,zhang2019hot,jiang2019dynamics}. That being said, the anti-symmetric part of the tensor and its effects on nonadiabatic dynamics have been much less understood \cite{bode2012current, teh2021antisymmetric,bajpai2020spintronics}. Previous studies have shown that the friction tensor can exhibit anti-symmetric terms in two situations. Firstly, Lu $et$ $al.$ \cite{lu2010blowing,lu2012current, bode2012current,lu2015current} found that a steady direct current via the presence of two leads with different Fermi levels (out of equilibrium) can make the anti-symmetric friction tensor non-zero, leading to a current-induced Berry force. Such a Berry force has significant effects on the current-induced dynamics \cite{lu2012current,erpenbeck2018current,ke2023current, preston2020current,kershaw2020non}. Secondly, Subotnik $et$ $al.$ \cite{subotnik2019demonstration} recently showed that spin-orbital couplings (SOC) or complex Hamiltonian can also lead to a Lorentz-like Berry force even at equilibrium \cite{teh2021antisymmetric,teh2022spin}. The Berry force arising from SOC is referred to as the spin Berry force. In both cases, the anti-symmetric part of the electronic friction tensor can contribute to a Lorentz-like Berry force. It was also noted that the Berry force can exist even without a continuum of electronic states. 

In our recent study \cite{mosallanejad2022floquet}, we investigated the photoinduced Lorentz-like Berry force using the Floquet theory, where we considered light as a periodic driving. We formulated a quantum classical Liouville equation (QCLE) in Floquet representation to describe nonadiabatic dynamics in open quantum systems with light-matter interactions (LMIs). We then mapped Floquet QCLE onto the EF model and found that the anti-symmetric friction tensor is no longer zero after introducing light (even without voltage bias), demonstrating that LMIs can independently give rise to a Lorentz-like Berry curvature force.
Moreover, we showed that the anti-symmetric friction can be larger than the symmetric one when the interaction between light and the system is relatively large. Therefore, the effects of the photoinduced Berry force on nuclear motion cannot be underestimated. However, understanding how photoinduced Berry curvature affects nonadiabatic dynamics exactly is the remaining question from our previous work, which will be further studied in the current paper. Answering this question will be very beneficial for photochemistry as well \cite{bernardi1996potential, sukharev2017optics, ebbesen2016hybrid, hertzog2019strong, honeychurch2023quantum, martinez2018can, george2015ultra,ibrahim2018h2}. 

In this paper, we study the joint influence of LMIs and bias voltage on Berry force, and their effects on nonadiabatic dynamics and quantum transport (including the kinetic energy and current). We organize this article as follows: in Sec. II, we introduce our model Hamiltonian and the Floquet EF model. We first show how to compute the Floquet EF model via non-equilibrium Green's functions; in Sec. III, we present and discuss our results on the anti-symmetric friction tensor, phonon relaxation dynamics, and localized current under different periodic drivings and biases. Finally, we conclude in Sec. IV. 
\section{Theory}
\label{sec:theory}
\subsection{A. Model Hamiltonian}
We consider a general Hamiltonian to describe the quantum transport of a molecular junction. The total Hamiltonian can be described as an electronic Hamiltonian plus the nuclear kinetic energy:
\begin{eqnarray}
\hat{H}_{tot} &=& \hat{H}_{el} +  \sum_{\mu} \frac{P_{\mu}^2}{2M_{\mu}}. 
\end{eqnarray}
The subscript $\mu$ indicates the nuclear degree of freedom (DoF). The electronic Hamiltonian $\hat{H}_{el}$ can be separated into three parts: the molecule $\hat H_s$, the leads $\hat H_b$, and the interactions between them $\hat H_c$: 
\begin{eqnarray}
\hat{H}_{el} &=& \hat H_s  + \hat H_b + \hat H_c,   \\
\hat H_s  &=& \sum_{ij} [h_s]_{ij} (\bold R, t) \hat d_i^\dagger \hat d_j + U(\bold R), \\
\hat H_b &=& \sum_{\zeta k} \epsilon_{\zeta  k} \hat c_{k\zeta}^\dagger \hat c_{k\zeta },  \\
\hat H_c  &=&  \sum_{\zeta k,i} V_{\zeta k,i}  (   \hat c_{k\zeta}^\dagger \hat d_i + \hat d^\dagger_i  \hat c_{k\zeta}  ). 
\end{eqnarray}
Here, $\hat d_j$ and $\hat c_{k\zeta}$ are annihilation operators for the level in the molecule and the leads, respectively. $U(\bold R)$ is the bare potential for the nuclei. We use $\zeta = L, R$ to denote left and right leads, and define the hybridization function $\Gamma_{\zeta ij}(\epsilon) = \sum_{ k} V_{\zeta k,i} V_{\zeta k,j} \delta (\epsilon-\epsilon_{\zeta k}) $ to characterize the system-bath couplings. 

\begin{figure}[ht]
\includegraphics[width=9cm]{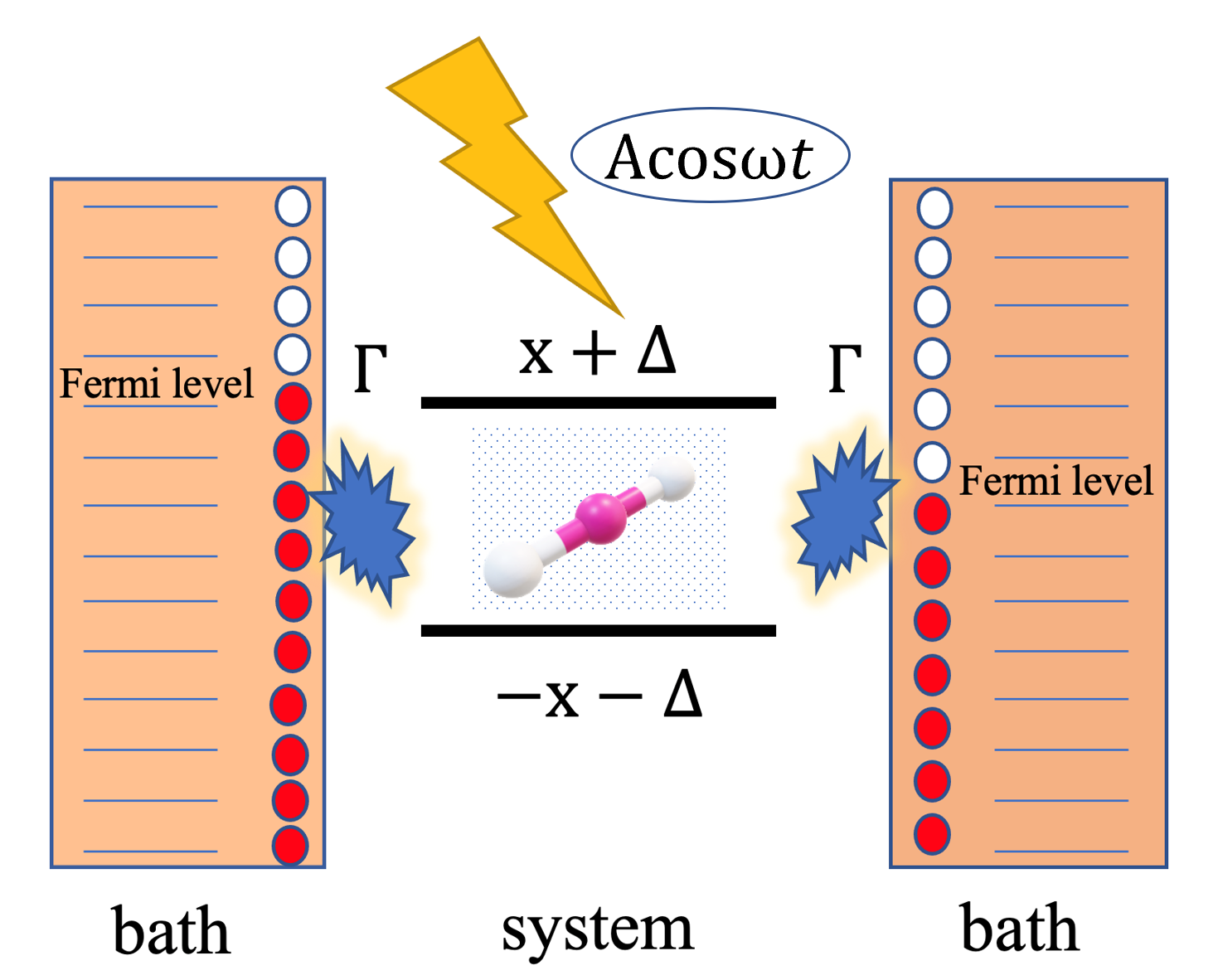}
\caption{A schematic picture for the model Hamiltonian: a two-level molecular junction is coupled with two baths (leads) with different Fermi levels. The light introduces couplings between the levels.}
\label{sk}
\end{figure}

Note that we have introduced light-matter interactions (LMIs) in the system Hamiltonian $\mathbf{h_s}(\mathbf{R},t)$. Here, we treat the light as a classical, time-dependent external field that couples to the system, such that the system Hamiltonian is time-periodic: $\mathbf{h_s}(\mathbf{R},t+T) = \mathbf{h_s}(\mathbf{R},t)$. 
Without loss of generality, we consider a two-level model with two-nuclear DoFs: 
\begin{eqnarray}
\mathbf{h_s}(\mathbf{R},t) = 
    \begin{pmatrix}
     x+ \Delta & y + A\cos(\omega t) \\
     y + A\cos(\omega t) & -x - \Delta
    \end{pmatrix},
\end{eqnarray}
where $x$ and $y$ are nuclear DoFs. $\Delta$ on the diagonal part of the Hamiltonian is the energy difference between the two levels, in analogy to the driving force in the case of the two shifted parabolas. 
The off-diagonal terms of the Hamiltonian indicate the couplings between the two levels, which consist of nuclear couplings $y$ as well as LMIs $A\cos(\omega t)$. 
$A\cos(\omega t)$ is a semi-classical representation of LMIs: $A$ represents the strength of the interaction and $\omega$ is the frequency of the light. A schematic figure for our model Hamiltonian is shown in Fig. \ref{sk}.
In addition, we have set the bare nuclear potential $U(\textbf{R})$ to be a two-dimensional harmonic oscillator:
\begin{eqnarray}
U(\mathbf{R}) = \frac{1}{2}x^2 + \lambda_x{x} + \frac{1}{2}y^2 +\lambda_y{y},
\end{eqnarray}
where we have chosen the frequencies of a two-dimensional harmonic oscillator to be $1$ in both $x$ and $y$ directions.  $\lambda_x$ and $\lambda_y$ shift the center of the Harmonic oscillator. As will be shown below, the bare potential does not affect the electronic friction but can change the dynamics overall.

\subsection{B. Floquet EF model: adiabatic force, friction tensors, and random force}
Utilizing the Floquet EF friction model, a semi-classical approach, we address the intricate dynamics of the system described in the previous subsection. The EF model is valid when the timescales of the electrons and lights are faster than the nuclear motion. Such that we can trace out all electronic DoFs with LMIs. The resulting equation of motion for the nuclei is a Langevin equation: 
\begin{eqnarray}
  M_\mu \ddot R_\mu = F_\mu^F - \sum_\nu \gamma_{\mu\nu}^F \dot R_\nu +  \delta F_\mu^F. 
\end{eqnarray}
Here $\mu$ and $\nu$ indicate nuclear DoFs, while $M$ and $R$ represent nuclear mass and position correspondingly. 
As for the three terms in the right hand side, $F_\mu^F$ is the mean force (adiabatic force), $\gamma_{\mu\nu}^F$ is the friction tensor (damping force), and $\delta F_{\mu}^F$ denotes a Markovian random force. The frictional and random forces represent the first-order correction to the Born-Oppenheimer approximation, i.e., the cooperated electron and light states are faster than the nuclear motion \cite{dou2015frictional,dou2017born}.

Specifically, the adiabatic force, friction tensor, and correlation function of the random force are given by:  
\begin{eqnarray}
\begin{aligned}
\gamma_{\mu\nu}^F = -& \frac{\hbar}{2\pi (2N+1)} \times \\
&\int_{-\infty}^{+\infty}d\epsilon \mathrm{Tr} \left \{ \partial_{\mu} h_s^F \partial_{\epsilon} G_r^F \partial_{\nu} h_s^F G_{<}^F \right \} + H.c.,\\
 F^F_{\mu}=  -&\frac{1}{2\pi i(2N+1)} \times \\
 &\int_{-\infty}^{+\infty}d\epsilon \mathrm{Tr}\left\{ \partial_{\mu} h_s^F  G_{<}^F \right\} - \partial_{\mu} U,\\
\frac{1}{2}(D_{\mu\nu}^F +& D_{\nu\mu}^F) =  \frac{\hbar}{4\pi (2N+1)} \times\\
&\int_{-\infty}^{+\infty}d\epsilon \mathrm{Tr}\left\{ \partial_\mu h_s^F  G_{>}^F \partial_\nu h_s^F  G_{<}^F \right\}.
\end{aligned}
\end{eqnarray}
Here, $Tr$ denotes trace over all electronic levels as well as Floquet levels. $n$ denotes the Floquet level ($n = -N \dots N $), such that we have $2N+1$ total Floquet states. $h_s^F$ is the Floquet representation of the system Hamiltonian: 
\begin{eqnarray}
h_s^F = \sum_{n = -N}^{N} h_s^{(n)}\hat{L}_n + \hat{N}\otimes\hat{I}_n \hbar \omega,\\
h_s^{(n)} = \frac{1}{T}\int_{0}^{T}h_s(R,t)e^{-in\omega t}dt,
\end{eqnarray}
where $\hat{L}_n$ and $\hat{N}$ are the ladder and number operators of the Floquet space. $\hat{I}_n $ is the identity operator of the Floquet space. $G_r^F$, $G_a^{F}$, $G_{<}^F$, and $G_{>}^F$ are the retarded, advanced, lesser, and larger Floquet Green’s functions of the electron in the energy domain, respectively. The Floquet Green's functions are given by:
\begin{eqnarray}
\begin{aligned}
\label{e19}
&G_r^{F}(\epsilon) = \left(  \epsilon-h_s^F -\Sigma_r^F  \right) ^{-1},\\
&G_a^{F}(\epsilon)=G_r^{F}(\epsilon)^\dagger,\\
&G^F_{<,>}(\epsilon)=G^F_{r} (\epsilon) \Sigma^F_{<,>} (\epsilon)   G^F_{a} (\epsilon).\\
\end{aligned}
\end{eqnarray}
Here $\Sigma_r^F$ is the retarded self-energy in Floquet space: $\Sigma_r^F = \Sigma_r \otimes \hat{I}_n $. We will apply the wide-band approximation such that $[\Sigma_r]_{ij} = -\frac{i}{2} \sum_\zeta \Gamma_{\zeta ij} $. $\Sigma^{F}_{<}$ (and $\Sigma^{F}_{>}$) is the lesser (and greater) self-energy in the Floquet space:  $ \Sigma^{F}_{ <} = \sum_\zeta \Sigma^{F}_{\zeta <} $, where $[\Sigma^{F}_{\zeta <}]_{ij}  = -i \Gamma_{\zeta ij} \otimes f(\epsilon - \hat{N}\hbar\omega -\mu_\zeta ) $.  

There are two levels in our model. We set orbital $1$ couples to the left lead and orbital $2$ couples to the right lead. Therefore, we can express the $\Gamma$ matrix as:
\begin{eqnarray}
\label{eq:gamma}
 \Gamma_L = 
    \begin{pmatrix}
      \widetilde{\Gamma} & 0 \\
      0  &  0
    \end{pmatrix},
 \Gamma_R = 
    \begin{pmatrix}
       0& 0 \\
      0  &  \widetilde{\Gamma}
    \end{pmatrix}.
\end{eqnarray}
Here, $\widetilde{\Gamma}$ is a constant within the wide-band approximation. We can define the hybridization function in the Floquet space as $\Gamma_L^F =  \Gamma_L  \otimes \hat{I}_n$ and similarly $\Gamma_R^F =  \Gamma_R  \otimes \hat{I}_n$.

Note that the friction tensor is not necessarily symmetric. In Ref. \cite{mosallanejad2022floquet}, we show that when introducing periodical driving, $(\gamma^F$ exhibits anti-symmetric part, which corresponds to a Lorentz-like force even at equilibrium (i.e., no electron voltage). Out of equilibrium, electron voltage can also introduce Lorentz-like forces. We expect the interplay of periodic driving and electronic voltage to give rise to an asymmetric friction tensor.

We get two random forces in both $x$ and $y$ directions relying on the correlation function of the random force ${D}_{\mu\nu}^F$, which is symmetric and positive-definite \cite{dou2017born, teh2021antisymmetric}:
\begin{eqnarray}
D^F = 
    \begin{pmatrix}
      D_{xx}^F &  D_{xy}^F \\
      D_{yx}^F &  D_{yy}^F
    \end{pmatrix}.
\end{eqnarray}
In our simulation, we generate two random numbers with a Gaussian distribution with norms $\sigma_\mu = \sqrt{2\widetilde{D}_\mu^F/dt}$. Here  $\widetilde{D}^F$ is the eigenvalue of the correlation function $D^F$. We then calculate the random force $\delta F_\mu^F$ by transforming the random numbers into the $x$ and $y$ basis.

\subsection{C. Evaluating electron current}
The total electron current is given by the integration of the local current over the phase space. In general, the local current with periodic driving is given by the Floquet Landauer formula \cite{kohler2005driven, haug2008quantum, meir1992landauer}:
\begin{eqnarray}
\begin{split}
    I_{loc}^F = &\frac{e}{2\pi\hbar(2N+1)} \times \\
    &\int_{-\infty}^{+\infty}d\epsilon Tr \left \{ T_{LR}^F(\epsilon) f_R^F(\epsilon)   - T_{RL}^F(\epsilon) f_L^F(\epsilon)\right \}.
\end{split}
\end{eqnarray}
Notice that the transmission probabilities are not necessarily equal to each other, $T_{LR}^F \neq T_{RL}^F$. For our symmetric coupling case, we can further simplify the formula:
\begin{eqnarray}
\begin{split}
I_{loc}^F = &\frac{e}{2\pi\hbar(2N+1)} \times \\
&\int_{-\infty}^{+\infty}d\epsilon Tr \left \{ T^F(\epsilon)[f_L^F(\epsilon)-f_R^F(\epsilon)] \right \}, 
\end{split}
\end{eqnarray}
where $f_L^F(\epsilon)$ and $f_R^F(\epsilon)$ are the Floquet Fermi-Dirac distributions for the left and right leads, and $T^F(\epsilon)$ is the Floquet transmission probability that can be expressed in terms of Green’s functions: 
\begin{eqnarray}
T^F(\epsilon) = \Gamma_L^F G^F_r(\epsilon) \Gamma_R^F G_a^F(\epsilon).
\end{eqnarray}

\section{Results and Discussions}
\label{sec:results}
\subsection{A. Anti-symmetric friction tensor}

\begin{figure*}[htbp]
    \centering
    \includegraphics[width=0.98\textwidth]{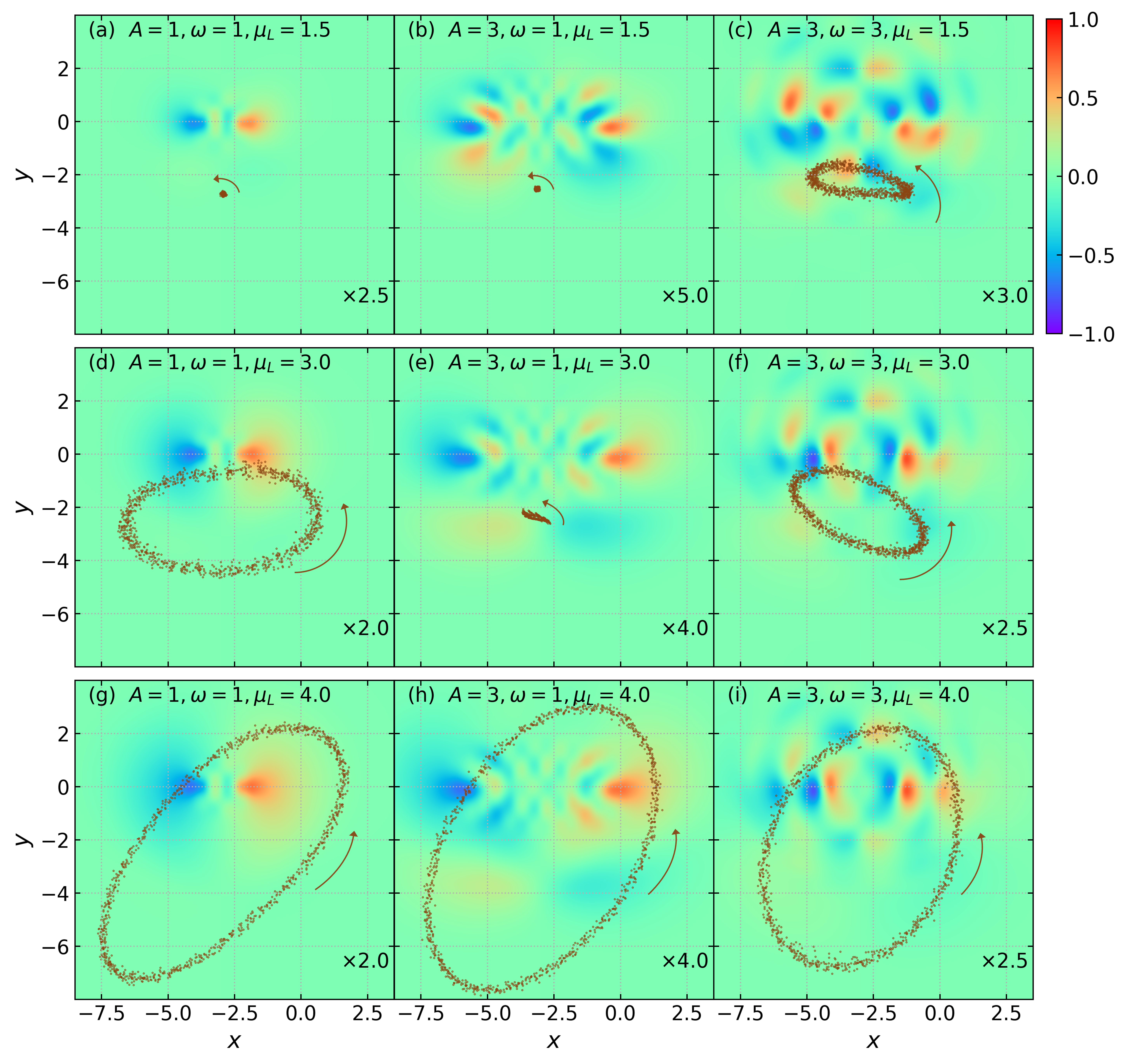}
    \caption{Anti-symmetric friction tensors in the presence of different driving and voltage bias and the corresponding stabilized trajectory of the nucleus without the fluctuating force.
    Parameters: $kT = 0.5$, $\Delta = 3$, $\mu_L = - \mu_R$, $\widetilde{\Gamma} = 1$, $\lambda_x = 3$, and $\lambda_y = 2$. We set the Floquet level $N = 5$ to ensure convergence. Note that the center of the potential is $(-\lambda_x, -\lambda_y)$, which is also the center the circular motion of the trajectories.}
    \label{fig:circle}
\end{figure*}

We first plot the anti-symmetric friction tensor as a function of the nuclear coordinates $(x, y)$ in Fig. \ref{fig:circle} for different driving frequencies and driving amplitudes. As mentioned above, in the absence of periodic driving and electron voltage, the friction tensor is symmetric and positive-definite. With periodic driving, the anti-symmetric part of the friction tensor exhibits peaks and dips in real space. As can be seen from each column of the figure, when increasing the bias, the shape of the anti-symmetric friction tensor does not change significantly, while the amplitude of the friction tensor increases with bias.
When we increase the driving amplitude, we see more peaks and dips, as shown in the figures. This is because when we have strong LMIs, more Floquet replicas are coupled in.  The overlaps of the Floquet replicas give rise to peaks and/or dips in the friction tensor. The positions of the peaks and dips are also very relevant to the driving frequency, as the driving frequency is related to the distance between the Floquet replicas. 

To illustrate the effects of the Lorentz-like force on the dynamics, we also plot the steady-state distribution of the trajectories in the same plots. Here, we disregard the random force in Langevin dynamics. 
Notice that the trajectories enter a cycle limit due to the Lorentz force. In general, with a large bias, we have a strong Lorentz-like force, such that the radius of the cycle is larger. The position of the circle is mainly determined by the mean force, which does not change significantly with the driving frequency and amplitude. The driving frequency and amplitude affect the nuclear distribution stringently, highlighting the importance of the Floquet modulation on nuclear dynamics.

\subsection{B. Langevin dynamics}
We now run the full Floquet Langevin dynamics with the fluctuating force, and plot the observable in the long time limit (cycle limit). In particular, we are interested in the kinetic energy as well as the electron current.

\begin{figure}[ht]
\includegraphics[width=8cm]{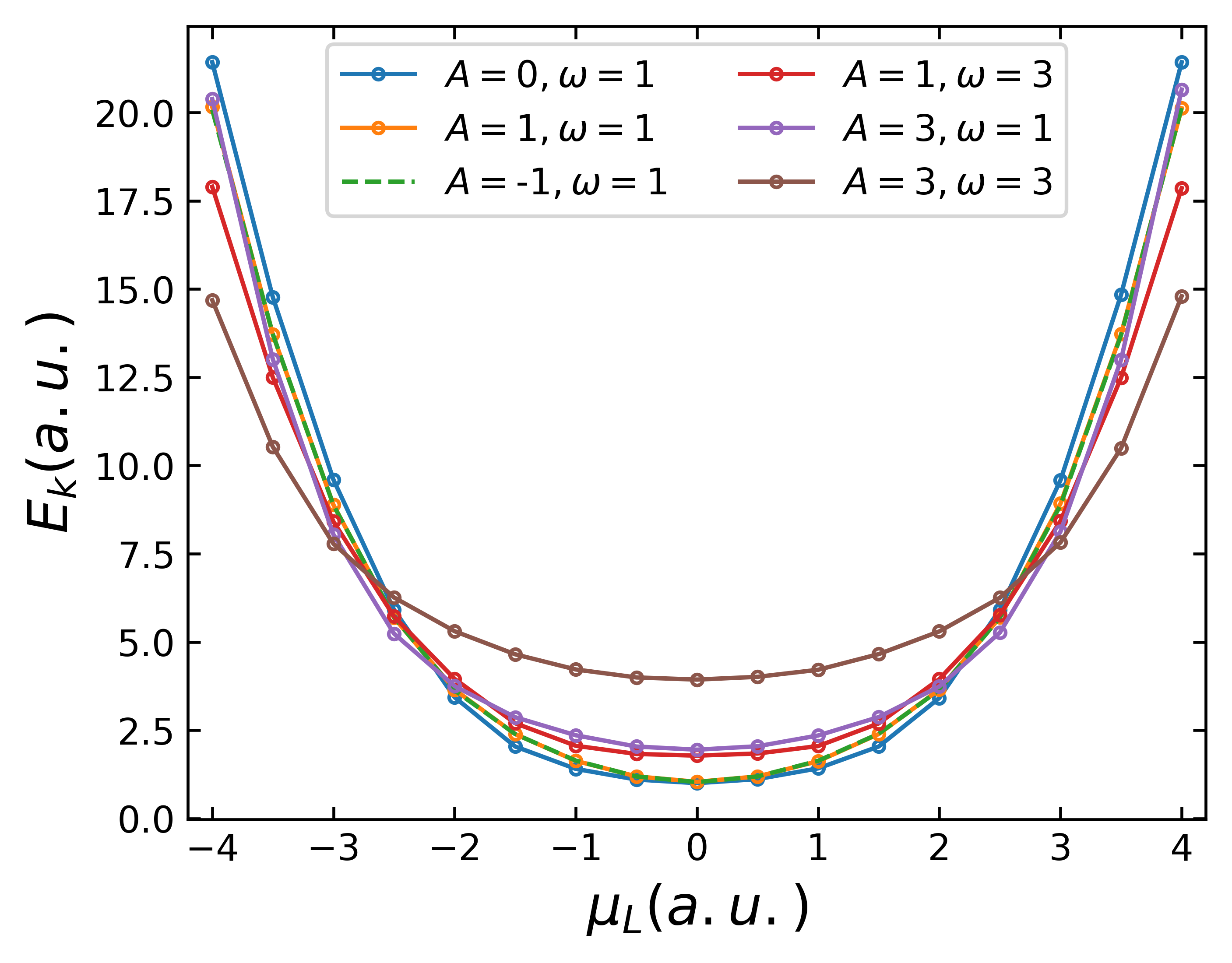}
\caption{Kinetic energy under different voltage bias and driving. 
Parameters: $kT = 0.5$, $\Delta = 3$, $\mu_L = - \mu_R$, $\widetilde{\Gamma} = 1$, $\lambda_x = 3$, and $\lambda_y = 2$.
$10,000$ trajectories are used to obtain the averaged kinetic energy}
\label{fig:kinetic}
\end{figure}

In Fig. \ref{fig:kinetic}, we plot the kinetic energy for the nuclei in the cycle limit for different voltage biases and driving amplitudes and frequencies. 
Remarkably, when the bias is $0$ and without any external driving ($A=0$), the total kinetic energy is equal to $kT$,
which constitutes the equilibrium kinetic energy baseline.
However, upon introducing an external driving ($A\neq 0$), the total kinetic energy surpasses the $kT$ threshold, indicating the emergence of heating effects induced by the external driving.
Furthermore, our observations reveal that the parameter $A$, whether set at $+1$ or $-1$, does not exert any discernible influence on the kinetic energy. This suggests that, within the context of our system, the direction of the incident light does not alter the thermal consequences.
Particularly, a more substantial driving amplitude corresponds to intensified heating.
Notice that the Lorentz force preserves the kinetic energy unchanged, while the second fluctuation and dissipation theorem becomes inapplicable in the presence of Floquet driving. 
Voltage bias can also introduce heating effects. 
That being said, the combined presence of Floquet driving and voltage bias doesn't invariably result in intensified heating effects.
Rather, at a larger bias (e.g., $\mu_L = 4$), the presence of the Floquet driving can diminish the heating effects. 
Conversely, external driving will most likely heat up the nuclei at low voltage bias.

\begin{figure}[ht]
\includegraphics[width=8cm]{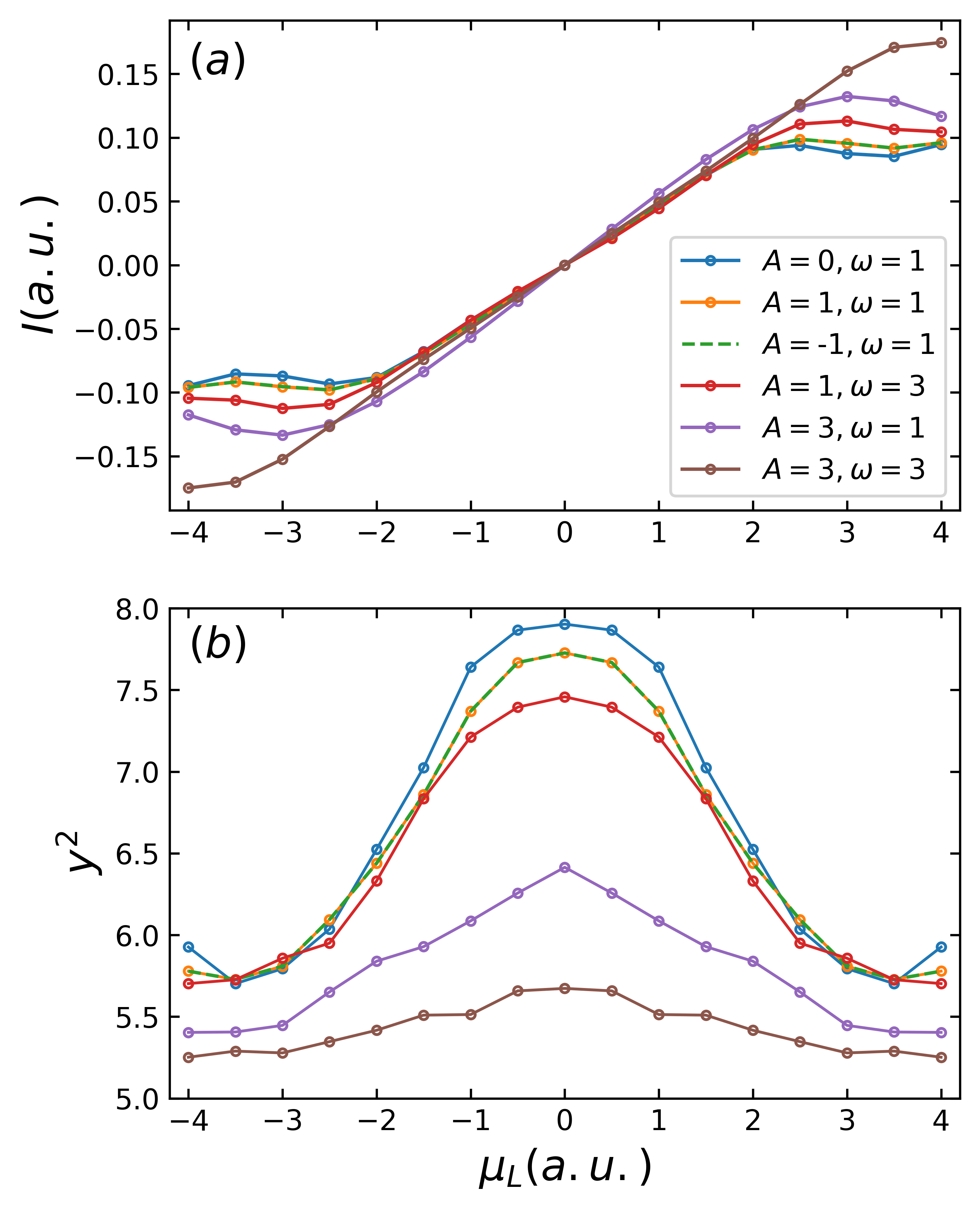}
\caption{Electron current under different bias and driving.
Parameters: $kT = 0.5$, $\Delta = 3$, $\mu_L = - \mu_R$, $\widetilde{\Gamma} = 1$, $\lambda_x = 3$, and $\lambda_y = 2$. 
The total current is averaged over $10,000$ trajectories. }
\label{fig: IV}
\end{figure}

Subsequently, we plot the total electron current under varying driving amplitudes and driving frequencies in Fig. \ref{fig: IV}. 
Note that without external driving, a larger bias can lead to a current profile exhibiting negative resistance characteristics.
Such a phenomenon stems from nuclear motion.
The total current significantly hinges on the off-diagonal couplings linking the two energy levels.
In addition, the off-diagonal coupling is position-dependent. 
In the lower panel of Fig. \ref{fig: IV}, we plot the averaged off-diagonal coupling between the two levels.
We compute the mean $y$-value for each trajectory upon entering the circular limit, followed by an averaging procedure across all trajectories. To accentuate distinctions among various couplings, we subsequently squared the resultant values.
Notice that increasing the bias can decrease the off-diagonal coupling, which can result in a decrease in the total electron current. 
In other words, the voltage bias affects the nuclear distribution, which decreases the off-diagonal coupling between two levels. 
Upon introducing external driving, the off-diagonal coupling's propensity to decrease with bias weakens, thereby mitigating the prominence of the negative resistance phenomenon.

\begin{figure}[ht] 
\includegraphics[width=8 cm]{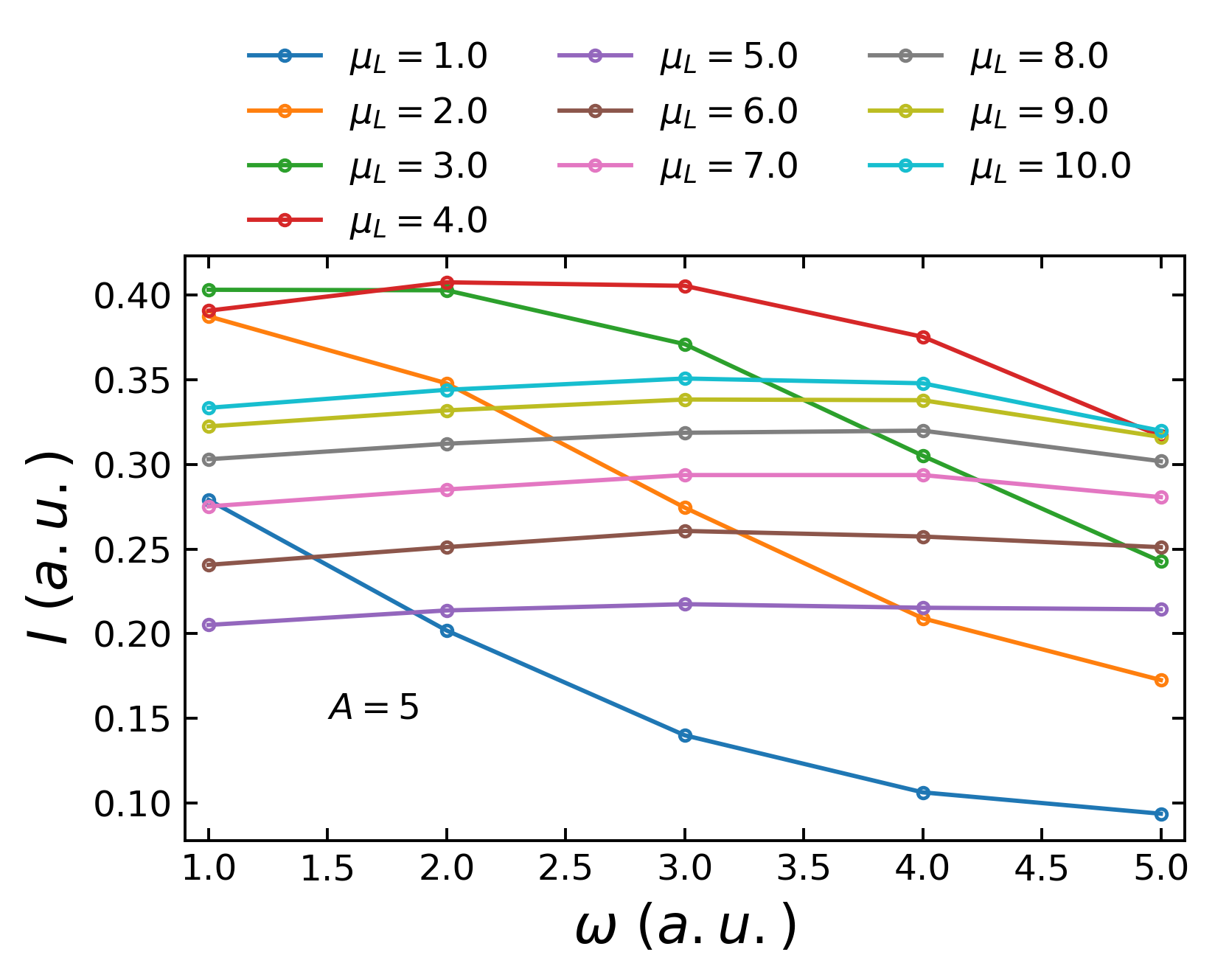}
\caption{Relationship between electron current and driving frequency in strong LMI regime $A = 5$.
Parameters: $kT = 0.5$, $\Delta = 3$, $\mu_L = - \mu_R$, $\widetilde{\Gamma} = 1$, $\lambda_x = 3$, and $\lambda_y = 2$.}
\label{fig:Iw}
\end{figure}

In Fig. \ref{fig:Iw}, we plot the total electron current as a function of the driving frequency. Here, we are operating within the strong LMI regime ($A=5$). When the bias is small, the current decreases with the driving frequency. For a larger bias, we see a turnover of the current as a function of driving frequency, i.e., there is an optimal frequency that maximizes the electronic current. That being said, the position of the optimal frequency highly depends on the voltage bias and driving amplitude.  

\section{Conclusions}
\label{sec:conclusion}
In summary, we have investigated the effects of the photoinduced Lorentz-like force on molecular motion in quantum transport. Our method is based on the Floquet electronic friction model, with the nonadiabatic dynamics of the coupled light and electronic motion being incorporated into a friction force and random force. We show that the anti-symmetric Lorentz force can redistribute the nuclear motion in the long time limit (cycle limit). In return, the nuclear motion affects electron transport strenuously. In particular, we observe that external driving can enhance and/or reduce electron current under different conditions.  Such that there is an optimal driving frequency that maximizes the electron current.  

Finally, we recall that Teh $et$ $al.$ recently reported SOC can also give rise to a Lorentz-like Berry force, and such a Berry force on nuclear motion amplifies the spin polarization of the electronic current significantly \cite{teh2021antisymmetric, teh2022spin}. Looking forward, we aim to investigate how LMIs affect the SOC-induced Berry force. How do the LMIs affect the spin current? Can we use light to control chiral induced spin selectivity? This work is ongoing.
\section{Acknowledgements}
We thank useful discussions with Hung-Hsuang Teh. W.L. thanks Zhixiang Dai and Siteng Ma for their help in the GPU calculation. This paper is based on work supported by the National Natural Science Foundation of China (NSFC Grant No. 22273075). We acknowledge the startup funding from Westlake University and the support of the high-performance computing center of Westlake University.

\bibliography{ref.bib}
\end{document}